\documentclass[twocolumn,english,aps,prl,showpacs]{revtex4}
\usepackage[T1]{fontenc}
\usepackage{amsmath,graphicx,amssymb,epsfig,babel,dsfont}

\renewcommand{\Im}{{\rm Im}}
\newcommand{\Tr}{{\rm Tr}}
\newcommand{\rd}{{\rm d}}
\newcommand{\kb}{k_{\rm B}}

\begin{document}

\title{Heat superdiffusion in plasmonic nanostructure networks}

\author{Philippe Ben-Abdallah}
\email{pba@institutoptique.fr}
\affiliation{Laboratoire Charles Fabry,UMR 8501, Institut d'Optique, CNRS, Universit\'{e} Paris-Sud 11,
2, Avenue Augustin Fresnel, 91127 Palaiseau Cedex, France.}
\author{Riccardo Messina}
\affiliation{Laboratoire Charles Fabry,UMR 8501, Institut d'Optique, CNRS, Universit\'{e} Paris-Sud 11,
2, Avenue Augustin Fresnel, 91127 Palaiseau Cedex, France.}
\author{Svend-Age Biehs}
\affiliation{Institut f\"{u}r Physik, Carl von Ossietzky Universit\"{a}t, D-26111 Oldenburg, Germany.}
\author{Maria Tschikin}
\affiliation{Institut f\"{u}r Physik, Carl von Ossietzky Universit\"{a}t, D-26111 Oldenburg, Germany.}
\author{Karl Joulain}
\affiliation{Institut PPrime, CNRS-Universit\'{e} de Poitiers-ENSMA, UPR 3346, 2, Rue Pierre Brousse, B.P. 633, 86022 Poitiers Cedex, France.}
\author{Carsten Henkel}
\affiliation{ Institut f\"{u}r Physik und Astronomie, Potsdam Universit\"{a}t, Germany}

\date{\today}

\pacs{44.05.+e, 12.20.-m, 44.40.+a, 78.67.-n}

\begin{abstract}
Heat-transport mechanism mediated by near-field interactions in plasmonic nanostructures networks is shown to be analogous to a generalized random-walk process. Existence of superdiffusive regimes is demonstrated both in linear ordered chains and in three dimensional random networks by analyzing the asymptotic behavior of the corresponding probability distribution function. We show that the spread of heat in these networks is described by a type of L\'{e}vy flight. The presence of such anomalous heat transport regimes in plasmonic networks opens the way to the design of a new generation of composite materials able to transport heat faster than the normal diffusion process in solids.
\end{abstract}

\maketitle

It is commonly admitted that heat conduction in massive solids is governed by a normal diffusion process (NDP). Heat carriers (phonons or electrons) move through the atomic lattice of material following a random walk \cite{Einstein} with a step length probability density which is a Gaussian. The speed of heat propagation and its spatial spreading from regions of high temperature to regions of low temperature is therefore intrinsically limited both by the speed of heat carriers and by the distance covered by them between two successive collision events. To go beyond this transport mechanism and accelerate the heat propagation within the medium we propose here to add a supplementary channel for heat exchanges with long-range interactions such as those that exist in generalized random walks (GRW), processes where the step length probability is broadband. L\'{e}vy flights \cite{Levy} are probably the most famous class of GRW in which extremely long jumps can occur as well as very short ones. Existence of photonic L\'{e}vy flight have been recently demonstrated \cite{Wiersma} in self-similar materials, the so called L\'{e}vy glasses. In those media, appropriately engineered so that photons perform random jumps with a probability distribution of step lengths which decays algebraically, the transport of propagative photons becomes superdiffusive. However, the magnitude of heat flux which can be transported with radiative photons is limited by the famous Stefan Boltzmann law \cite{Planck} and is several orders of magnitude smaller than the flux carried by conduction in solids.
The situation radically changes when these photons become non-radiative. As predicted by Polder and van Hove \cite{Polder1973} fourty years ago and experimentally verified during the last few years \cite{HuEtAl2008,ShenEtAl2008,RousseauEtAl2009,Ottens2011}, when two media out of thermal equilibrium are separated by a small distance (compared with their thermal wavelength) they exchange energy mainly by photon tunneling. In such a situation, the heat flux transported from one medium to the other one can surpass by several orders of magnitude the flux exchanged between two blackbodies \cite{VolokitinPersson07,Joulain} in far field. In two recent works \cite{PBA2011PRL,Messina} we have established that a similar exalted heat transport can also exist at larger distances thanks to many-body interactions. In this Letter we investigate in detail how heat is transported throughout different plasmonic nanostructure networks which are either ordered or disordered. By analyzing the transport process through these structures as a GRW of a passive tracer in a medium, we demonstrate the existence of anomalous (superdiffusive) regimes driven by the collective near-field interactions.

To start this anlalysis, let us consider an arbitrary three-dimensional network of spherical particles of radius $R_i$ at temperature $T_i$ distributed inside an environment at temperature $T_\text{env}$. When the mean separation distance between two arbitrary particles is larger than their respective diameter and their size is small enough compared with the thermal wavelengths $\lambda_{T_i}=c\hbar/(\kb T_i)$ then this network can be modelized by a set of simple electric and magnetic dipoles in mutual interaction and in interaction with the surrounding bath. The time evolution of particle temperatures in presence of external excitation is governed by the following energy balance equation
\begin{equation}
\rho_i C_iV_i\frac{\partial T_i}{\partial t}=\underset{j\neq i}{\sum}\mathcal{P}_{i\leftrightarrow j}+\mathcal{P}_{i\leftrightarrow B}+S_i\label{Eq:heat_eq},
\end{equation}
where $\rho_i$ and $C_i$ represent the nanoparticle mass density and heat capacity respectively while $\mathcal{P}_{i\leftrightarrow j}$, $\mathcal{P}_{i\leftrightarrow B}$ and $S_i$ denote the net power exchanged between two arbitrary particles, the power exchanged between a particle and the thermal bath and the power received by a particle from an external source respectively. Using the Landauer formalism introduced in \cite{JoulainPBA2010,BiehsEtAl2010} and extended in \cite{PBA2011PRL} to the $N$-body heat-transport problem, it can be shown that
\begin{equation}
\mathcal{P}_{i\leftrightarrow j}=3\int_{0}^{\infty}\frac{\rd\omega}{2\pi}\,[\Theta(\omega,T_i)-\Theta(\omega,T_{j})]\mathcal{T}_{i,j}(\omega),\label{Eq:power}
\end{equation}
where
\begin{equation}
\mathcal{T}_{i,j}(\omega)=\frac{4}{3}\frac{\omega^{4}}{c^{4}}\Im(\alpha_i)\Im(\alpha_{j})\Tr\bigl[\mathds{G}_{ij}\mathds{G}^{\dagger}_{ij}\bigr]\label{Eq:transmission}
\end{equation}
denotes the monochromatic transmission coefficient between the dipoles $i$ and $j$ expressed in terms of the dyadic Green tensor $\mathds{G}_{ij}$ between the location of two dipoles and the particle polarizability $\alpha$ while $\Theta(\omega,T)=\hbar\omega/\bigl[\exp(\frac{\hbar\omega}{k_\text{B}T})-1\bigr]$ is the mean energy of a harmonic oscillator at temperature $T$. The electric polarizability is given by the simple Clausius-Mossotti form $\alpha=4\pi R^{3}\frac{\varepsilon-1}{\varepsilon+2}$. We have checked that the radiative correction \cite{Albaladejo} to this polarizability, which is proportional to $(k_0 R)^3$, is negligible for nanoparticles in the Wien frequency range under consideration. For calculating the dyadic Green tensor (GF) in a system of $M$ particles we use the coupled-dipole equation \cite{Purcell}
\begin{equation}
\mathbf{E}_{ij}=\mu_{0}\omega^2\mathds{G}_{0}^{ij}\mathbf{p}^\text{fluc}_{j\neq i}+\frac{\omega^2}{c^2}\underset{k\neq i}{\sum}\mathds{G}_{0}^{ik}\alpha_k\mathbf{E}_{kj},\label{Eq:Lax}\end{equation}
for $i=1,...,M$, combined with the linear response of a dipole source
\begin{equation}
\mathbf{E}_{ij}=\omega^2\mu_{0}\mathds{G}_{ij}\mathbf{p}^\text{fluc}_{j}.
\label{Eq:GreensFunction}
\end{equation}
Eq. \eqref{Eq:Lax} gives the field exciting the particle $i$ and coming from particle $j$. It contains a direct contribution, associated to the fluctuating dipole source $\mathbf{p}^\text{fluc}_{j}$, and the contributions induced by all other particles. Here $\mathds{G}_0^{ij}=\frac{\exp(\text{i}kr_{ij})}{4\pi r_{ij}}\left[\left(1+\frac{\text{i}kr_{ij}-1}{k^2r_{ij}^2}\right)\mathds{1}+\frac{3-3\text{i}kr_{ij}-k^2r_{ij}^2}{k^2r_{ij}^2}\widehat{\mathbf{r}}_{ij}\otimes\widehat{\mathbf{r}}_{ij}\right]$
is the free space GF defined with the unit vector $\widehat{\mathbf{r}}_{ij}\equiv\mathbf{r_{\mathit{ij}}}/r_{ij}$, $\mathbf{r_{\mathit{ij}}}$ being the vector linking the center of dipoles $i$ and $j$, while $r_{ij}=|\mathbf{r}_{ij}|$ and $\mathds{1}$ stands for the unit dyadic tensor. The second term of the rhs of Eq. \eqref{Eq:heat_eq} is the power exchanged in far field with the environnement and it can be expresses as
\begin{equation}
\mathcal{P}_{i\leftrightarrow B}=\overline{C}_{\text{abs};i}\sigma_{B}(T_\text{env}^{4}-T_i^{4})\label{Eq:heat_far_field},
\end{equation}
where $\overline{C}_{\text{abs};i}$ is the thermally averaged dressed absorption cross-section of the $i$-th particle \cite{Yannopapas}. At the beginning of the thermal relaxation process, the power exchanged between the particles through near-field interactions are much more important than the power exchanged in far field with the environnement so that $\mathcal{P}_{i\leftrightarrow B}$ can be neglected with respect to the other terms in Eq. \eqref{Eq:heat_eq}. Moreover, in a situation of local quasi-thermal equilibrium the net power exchanged between two dipoles reads, according to Eq. \eqref{Eq:power},
\begin{equation}
\mathcal{P}_{i\leftrightarrow j}=G(|\mathbf{r}_i-\mathbf{r}_j|;T_i)(T_j-T_i)\label{Eq:power_linéarisé}
\end{equation}
where we have introduced the thermal conductance at temperature $T_i$ between the points $i$ and $j$
\begin{equation}\begin{split}
G(|\mathbf{r}_i-\mathbf{r}_j|;T_i)&\equiv\frac{\partial\mathcal{P}_{i\leftrightarrow j}}{\partial T}{(T_i)}\\
&=3\int_{0}^{\infty}\frac{\rd\omega}{2\pi}\,\frac{\partial\Theta(\omega,T_i)}{\partial T}\mathcal{T}_{i,j}(\omega).\end{split}\label{Eq:conductance}\end{equation}

Using this expression, the energy balance equation \eqref{Eq:heat_eq} can be recast into a Chapman-Kolmogorov master equation
\begin{equation}
\frac{\partial T_i}{\partial t}=\intop_{R^{d}}p(\mathbf{r}_i,\mathbf{r})\frac{T(\mathbf{r},t)}{\tau(\mathbf{r})}d\mathbf{r}-\frac {T(\mathbf{r}_i,t)}{\tau(\mathbf{r}_i)}+\hat{S}_i \label{Eq:heat_eq4},
\end{equation}
which formally describes a system which is driven by a Markov process. $T(\mathbf{r},t)$ plays here the role of a passive scalar which is subject to transitions or random jumps that modify its state between two positions $\mathbf{r}$ and $\mathbf{r}'$ with a probability density (not normalized) function (pdf) $p(\mathbf{r},\mathbf{r^{'}})=\frac{1}{\rho C V\Delta{V}}\tau(\mathbf{r})G(\mathbf{r}-\mathbf{r}')$ at a time rate of $\tau$ with $\tau^{-1}(\mathbf{r})=\frac{1}{\rho C V\Delta{V}}\intop_{R^{d}}G(\mathbf{r}-\mathbf{r}')d\mathbf{r}'$ in presence of a localized source $\hat{S}(\mathbf{r})=S(\mathbf{r})/(\rho C V)$ [$d$ denotes here the space dimension where heat exchanges occur and $\Delta {V}$ is an elementary volume defines with the average distance $\overline{l}$ between two neighboring dipoles inside the system as $\Delta{V}=\overline{l}^{3}$]. To analyze the transport of heat throughout a given plasmonic network we can investigate how the pdf decays with respect to the distance. When this decreasing is Gaussian, any moment $M^{(n)}=\int x^{n}p(x,t)dx$ of the pdf is finite so that the regime of transport is diffusive. On the other hand if it decays algebraically, so that at least one step moment is divergent, the heat transport regime becomes superdiffusive.

\begin{figure}[Hhbt]
\includegraphics[scale=0.35]{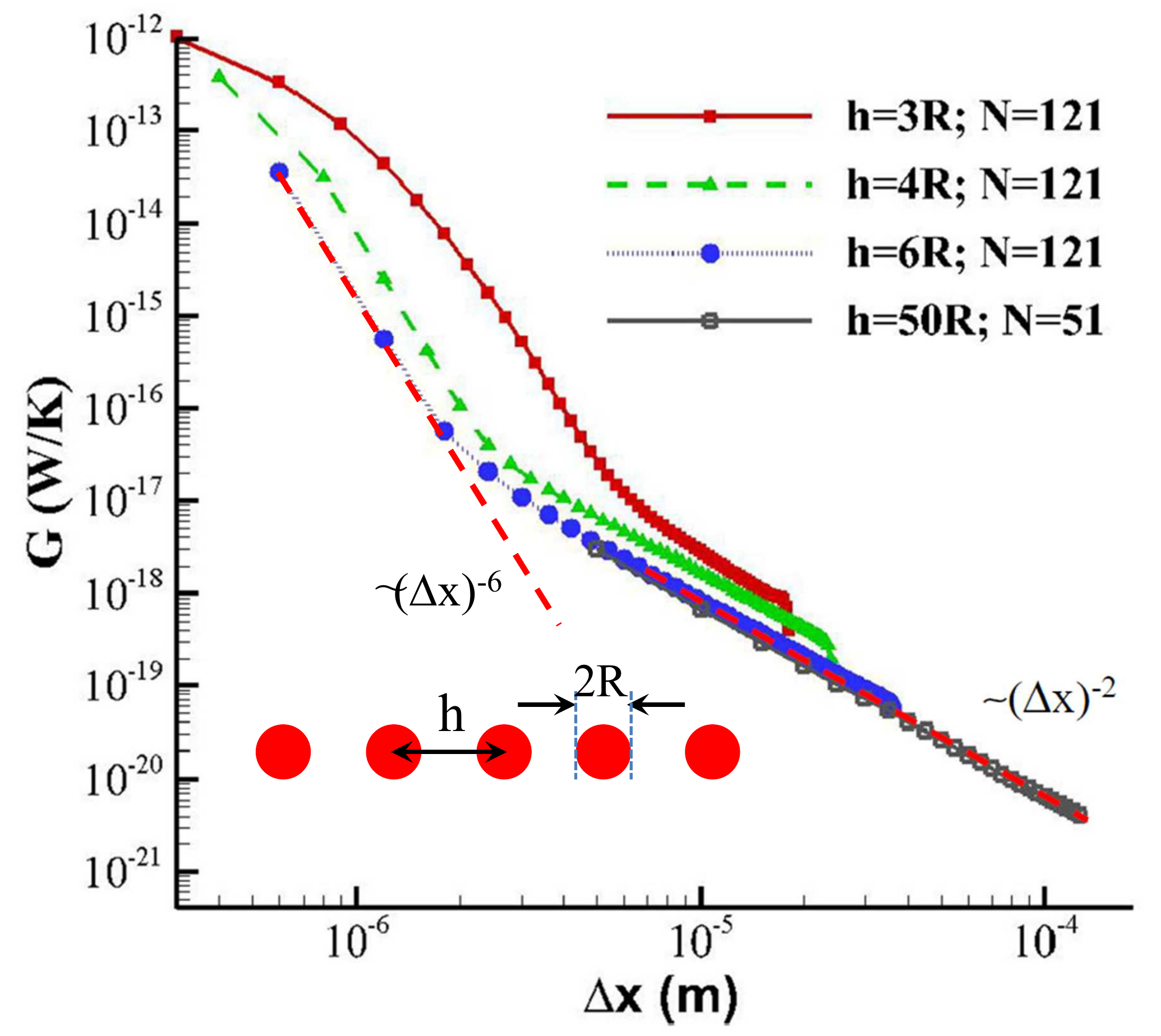}
\caption{Thermal conductance $G$ in log-log scale for a chain of SiC spherical particles with different inter-particle distances $h$ and different particle numbers $N$ as a function of the separation distance $\Delta x$ at temperature $T=300\,$K. All particles are identical (radius of $R=100\,$nm) and their electric polarizability is given by the simple Clausius-Mossotti form $\alpha=4\pi R^{3}\frac{\epsilon-1}{\epsilon+2}$ \cite{Albaladejo}. The dielectric permittivity of the particles is described by a Drude-Lorentz model \cite{Palik}. The end of curves is slightly curved due to boundary effects.}
\end{figure}

We first consider the case of linear chains ($d=1$) of nanoparticles periodically dispersed in a transparent medium. The thermal conductance between two particles, inside a given chain, is plotted in Fig. 1 for different inter-particle separation distances $h$. We observe, for any value of $h$, that at large separation distance $x$ the tail of $G(x)$ asymptoticaly decays as $G\sim O(\frac{1}{x^2})$.
By neglecting the boundary effects and remarking first that $\tau$ is almost uniform (i.e. $\tau(x)\sim\tau_0$) and then that $p(\mathbf{x},\mathbf{x'})$ is a function of $x-x'$ only, the spatial Fourier transformation of Eq. \eqref{Eq:heat_eq4} gives
\begin{equation}
\frac{\partial \tilde{T}(k,t)}{\partial t}=\frac{1}{\tau_0}[ \tilde{p}(k)-1]\tilde{T}(k,t)+\tilde{\hat{S}}_i \label{Eq:heat_Fourier}.
\end{equation}
In the small wavenumber approximation (i.e. hydrodynamic limit) we get, after developing $\tilde p(k)$ around $k=0$ and coming back to the real space, the 1D Fokker-Planck equation (FPE) which describes the transport process
\begin{equation}
\frac{\partial {T}(x,t)}{\partial t}=A T(x,t)-V\frac{\partial {T}(x,t)}{\partial x}+D \frac{\partial^2 {T}(x,t)}{\partial^2 x}+ \hat{S}\label{Eq:FPE},
\end{equation}
where $A=\frac{1}{\tau_0}[\tilde {p}(0)-1]$, $V=-\frac{1}{\tau_0} \frac{d\tilde {p}(0)}{dk}$ and $D=\frac{1}{\tau_0} \frac{d^2\tilde {p}(0)}{dk^2}$ denote the transport coefficient, the dynamic friction and the diffusion coefficient respectively inside the chains. By noting that $D$ is proportional to the moment of second order of the pdf, we immediately see according to the asymptotic behavior of the conductance tail (see Fig. 1) that the diffusion coefficient is diverging demonstrating so the superdiffusive behavior of heat transport in 1D dipolar chains. Besides, we clearly observe on these curves the transition between the region where the electrostatic regime ($\Delta x<<\lambda_T$) dominates and the region where interactions take place at distances longer than the wavelength. In the first case we distinguish two different behaviors. When $h=6R$ the thermal conductance follows a power law in $G\sim (\Delta x)^{-6}$ analog to what is usually observed between two isolated dipoles \cite{Domingues2005,Perez2008}. In denser chains the dependence of the conductance on the inter-particle distance is no longer the same. We see even in Fig. 1 that the thermal conductance tends to saturate at close separation distances in chains where extreme near-field interactions take place (i.e. $h=3R$). This behavior could be at the origin of the saturation mechanism that has been observed \cite{Kittel} in near-field for heat exchanges due to the non-local optical response of materials. However, today this problem still remains largely unexplored. At long distances (compared with $\lambda_T$) we see that $G\sim O(\frac{1}{x^2}$) for any chain. In the diluted chain ($h=50R$ in Fig. 1) all dipoles can be considered as isolated and they exchange heat in far field mainly with their nearest neighbors. As for the field magnitude (and that of dyadic Green's tensor), it evolves from each dipole as $1/x$, so that the thermal conductance follows a $1/x^2$ power law. In denser chains (typically when $h<6R$) collective effects \cite{PBA2008PRB} start playing a significant role in the heat transport along the chain. However at mesoscopic scale we recover the dependence in $1/x^2$ for the thermal conductance.

Now let us discuss the heat-transport process mediated by the near-field interactions in three-dimensional disordered networks made of $N$ identical nanoparticles of identical radius $R$ randomly distributed within a fictious cubic box of side $a$ as depicted in the inset of Fig. 2. Each realization is generated with a uniform distribution probability and a minimum distance $r_\text{min}\sim2R$ is imposed between two adjacent particles in order to keep the dipolar approximation valid. Moreover, in any generated realization, a nanoparticle $P_0$ occupies the center of the simulation box and is used as a reference particle for the calculation of thermal conductance. To avoid the presence of possible boundary effects which result from the finite size of the simulation box, we have limited our conductance calculations to spatial domains contained within spheres of radius $\widetilde{R}=\frac{1}{4}a$ and checked a posteriori this assumption. The corresponding values $G^{(i)}(r;T)$ of conductance for the $i$-th realization at a distance $r$ from $P_{0}$ are calculated by making an averaging over all particles located inside a shell of radius $r$ and of thickness $\delta\ll\widetilde{R}$. The average conductance $\langle G\rangle=\frac{1}{m}\underset{i=1}{\overset{m}{\sum}}G^{(i)}$ over the $m$ realizations is plotted in Fig. 2 with respect to the separation distance from $P_{0}$ for different volumic fractions $f=\frac{NV}{a^3}$. Inspection of Fig. 2 shows that $\langle G(x)\rangle$ decays in power law as $\zeta/x^{\gamma}$ with an exponent $\gamma$ which depends only on the filling factor $f$. According to this, the energy balance equation \eqref{Eq:heat_eq4} (its statistical averaging) can be recast into a fractional-like diffusion equation
\begin{equation}
\rho_i C_iV_i\frac{\partial T_i}{\partial t}=-\kappa(-\Delta)^{\alpha/2}T(\mathbf{r}_i)+S_i\label{Eq:heat_eq2},
\end{equation}
with $\alpha=\gamma-d$, for $\alpha\in[0,2]$. $(-\Delta)^{\alpha/2}$ denotes here the fractional Laplacian defined by \cite{Podlubny,Samko}
\begin{equation}
 (-\Delta)^{\alpha/2}T(\mathbf{r})=c_{d;\alpha}\,\text{P}\intop_{R^{d}}\frac{T(\mathbf{r})-T(\mathbf{r'})}{|\mathbf{r}-\mathbf{r'}|^{d+\alpha}}d\mathbf{r'}
\end{equation}
with $c_{d;\alpha}=\frac{2^{-\alpha} \pi^{1+d/2}}{\Gamma(1+\alpha/2)\Gamma(\frac{d+\alpha}{2})\sin(\alpha \pi/2)}$ and where P denotes the principal part. In Eq. \eqref{Eq:heat_eq2} $\kappa=\frac{\zeta}{\Delta V c_{d;\alpha}}$ is the fractional diffusion coefficient inside the plasmonic structure. The fractional Laplacian clearly shows that the smaller is $\alpha$, the larger is the length of interactions through the medium. When $\alpha=2$ the fractionnal Laplacian degenerates into a classical Laplacian and the heat transport process becomes diffusive. If $\alpha>2$ the transport is subdiffusive while for $\alpha<2$ it is superdiffusive. Contrary to NDPs, the long-ranged (non-local) interactions through the network is responsible for the existence of anomalous heat-transport regimes. At low $f$, the exponent $\gamma$ is close to 5 (i.e. $\alpha$ is close to $2$ but still smaller). However, even if the transport process tends toward a diffusive process, it still remains superdiffusive at this level of filling factor. This regime corresponds to plasmonic networks where the mean separation distances between the nearest particles is $\bar{l}>6.6R$. In this situation heat exchanges are not limited to the closest neighbors and collective effects continue to play a role despite the fact that the medium is quite diluted. At higher densities, $\alpha$ decreases showing that the heat-transport process is more and more non-local. At $f=20\%$ that is in networks where extreme near-field interactions occur ($\bar {l}\simeq 2.7R$) we have $\alpha\simeq 0.64$ so that the heat transport becomes unambiguously superdiffusive.

\begin{figure}[Hhbt]
\includegraphics[scale=0.5]{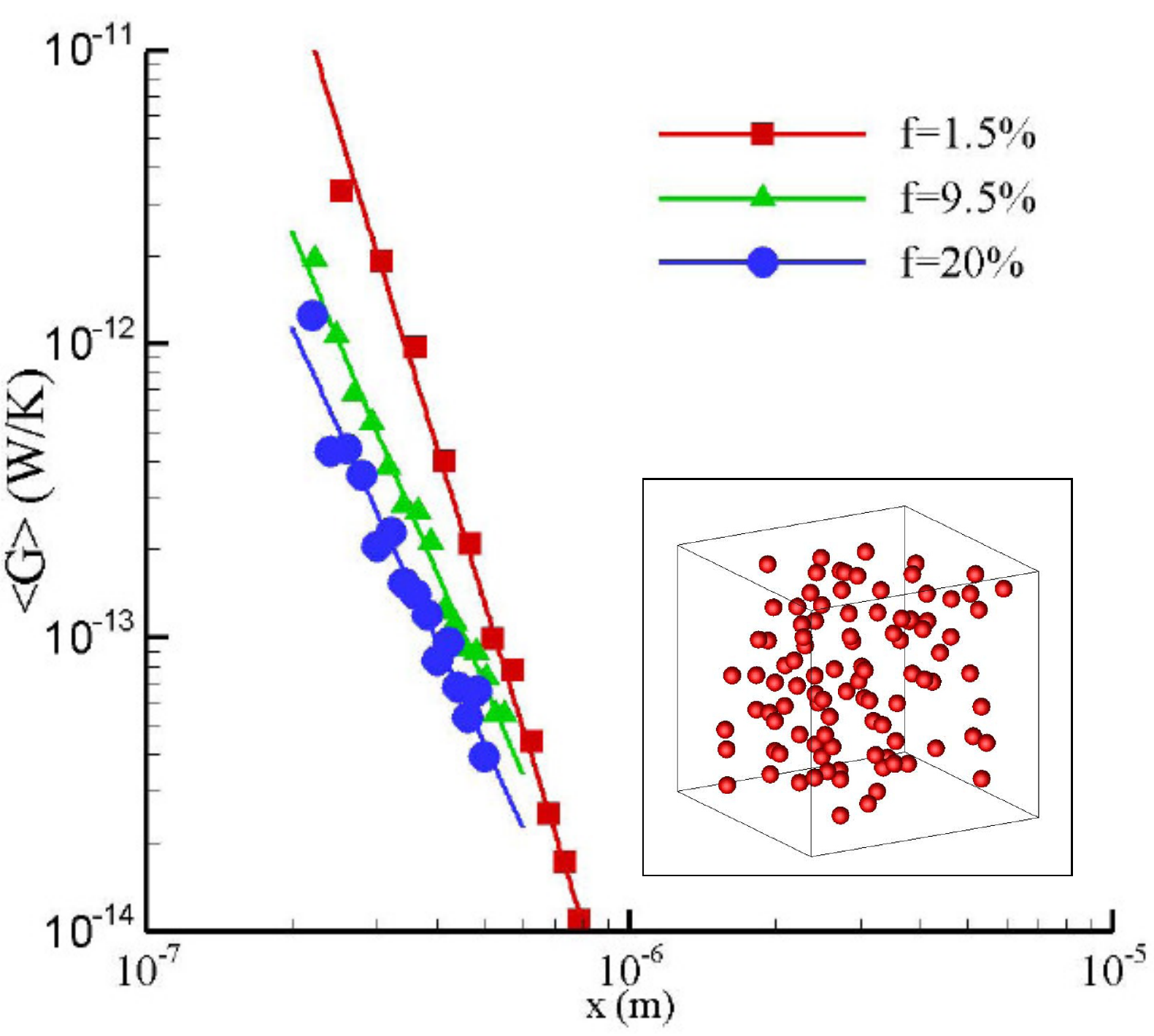}
\caption{Averaged thermal conductance $\langle G\rangle$ in log-log scale for clusters of SiC spherical particles as a function of the separation distance $x$ for different volumic fractions $f$ and at temperature $T=300\,$K. The statistical averaging is performed with $m=250$ realizations generated with a uniform random distribution probability. The inset shows an example of network with a volumic fraction $f=1.5\%$ generated with $N=100$ nanoparticles.}
\end{figure}

We have demonstrated that the heat transport mediated by photon tunneling in plasmonic networks can be extremely superdiffusive. This non-local heat-transport mechanism allows to go beyond the standard diffusion limit in solids. The ability to design nanocomposite materials able to transport heat faster than with phonons in solids opens new perspectives. It could find broad applications in different fields of material sciences that require an ultrafast thermal management. Many fascinating questions on the links between the spatial structuration of plasmonic structures networks and the transport of heat through them remain open. For instance, the role played by the disorder and the presence of localized and delocalized modes is one of them. Also, the phonon-photon coupling within the plasmonic structures embedded in solids is a fundamental issue because it affects the transition between the superdiffusive regime and the classical diffusive transport.

%
%

\begin{acknowledgments}
This work has been partially supported by the Agence Nationale de la Recherche through the
Source-TPV project ANR 2010 BLANC 0928 01. P.B.-A. thanks fruitfull discussions with P. Lalanne, H. Benisty and M. Rubi.
\end{acknowledgments}

\end{document}